\documentclass[12pt]{article}
\usepackage{amsmath,amssymb,mathrsfs,url,graphicx, color}
\usepackage{cite}

\date{}

\title{Time reversal invariance and ontology}

\author{ Ward Struyve\footnote{Centrum voor Logica en Filosofie van de Wetenschappen, KU Leuven, Belgium} \footnote{Instituut voor Theoretische Fysica, KU Leuven, Belgium}  }

\def\pa{\partial}

\def\al{\alpha}

\def\ii{\textrm i}
\def\ee{\textrm e}

\newcommand{\be}{\begin{equation}}
\newcommand{\en}{\end{equation}}
\newcommand{\trie}{time reversal invariance}
\newcommand{\trit}{time reversal invariant}

\begin{document} 
\maketitle

\begin{abstract}
\noindent
Albert and Callender have challenged the received view that theories like classical electrodynamics and non-relativistic quantum mechanics are time reversal invariant. They claim that time reversal should correspond to the mere reversal of the temporal order of the instantaneous states, without any accompanying change of the instantaneous state as in the standard view. As such, Albert and Callender claim, these theories are not time reversal invariant. The view of Albert and Callender has been much criticized, with many philosophers arguing that time reversal may correspond to more than the reversal of the temporal order. In this paper, we will not so much engage with that aspect of the debate, but rather deflate the disagreement by exploiting the ontological underdetermination. Namely, it will be argued that with a suitable choice of ontology, these theories are in fact time reversal invariant in the sense of Albert and Callender, in agreement with the standard view. 
\end{abstract}

\section{Introduction}
Physics textbooks state that theories like Newtonian mechanics, classical electrodynamics and non-relativistic quantum mechanics are time reversal invariant. Albert \cite{albert00} and Callender \cite{callender00} disagree. Albert claims that only the former is time reversal invariant, while the other two are not \cite[p.\ 14]{albert00}:
\begin{quote}
And so [classical electrodynamics] is not invariant under time reversal. Period.

And neither (it turns out) is quantum mechanics, and neither is relativistic quantum field theory, and neither is general relativity, and neither is supergravity, and neither is supersymmetric quantum string theory, and neither (for that matter) are any of the candidates for a fundamental theory that anybody has taken seriously since Newton. And everything everybody has always said to the contrary \dots\ is wrong.
\end{quote}
Callender discusses just non-relativistic quantum mechanics \cite{callender00}, but also arrives at the conclusion --- for the same reason as Albert --- that this theory is not time reversal invariant.

To explain the disagreement, let us consider Albert, who gives a detailed discussion of what he takes time reversal to mean. Consider first Newtonian mechanics. In this case, Albert actually agrees with the standard conclusion that the theory is \trit, but for different reasons. For Albert, the collection of positions of the particles at a time forms the instantaneous state. The temporal sequence of these instantaneous states forms a history. Albert takes the time reversal of a history to be just the history run backwards. That is, the temporal order of instantaneous states is reversed. It is as if a video of the motion of the particles is played backwards. Newtonian mechanics is \trit\ because the time-reversed of each dynamically allowed history is also dynamically allowed. That is, time reversal is a symmetry by turning solutions to Newton's dynamics into solutions. In the example of the video, the time reversal invariance entails that we would not be able to tell on the basis of Newton's dynamics whether the video is played backwards or not, since both evolutions are allowed by the dynamics.

Also according to the standard view Newtonian mechanics is time reversal invariant, but the story is a bit different. First of all, in addition to the particle positions, also the instantaneous velocities are included in the instantaneous state. In this way, the instantaneous state determines a unique solution to the Newtonian dynamics. Second, according to the standard view, the time reversal amounts to reversing the temporal order of the instantaneous states together with flipping the sign of the velocities at each time. So the time reversal is more than just reversing the temporal order of the instantaneous states. For Albert, the velocities also flip sign under time reversal, but that is because they are the rates of change of the positions and hence a time reversal of the positions induces a sign flip of the velocities. In any case, despite these differences, the conclusion is the same: Newtonian mechanics is \trit.

Disagreement arises in the case of classical electrodynamics. In this case, the electric and magnetic field are included in the instantaneous state, both according to Albert and standard textbooks. So there is no disagreement concerning the electromagnetic part of the instantaneous state. However, according to the standard view, the magnetic field flips sign under time reversal, like the velocities in Newtonian mechanics. But for Albert the magnetic field should not change sign under time reversal \cite[p.\ 20]{albert00}:
\begin{quote}
Magnetic fields are not the sorts of things that any proper time reversal transformation can possibly turn around. Magnetic fields are not---either logically or conceptually---the rates of change of anything. 
\end{quote}
As such Albert concludes that electrodynamics is not \trit.

The story in non-relativistic quantum mechanics is similar. There is no disagreement that the instantaneous state is given by the wave function. But standard time reversal involves more than merely reversing the temporal order of states, namely it also involves taking the complex conjugate of the wave function. Merely reversing the temporal order does not correspond to a symmetry of the theory and so Albert and Callender conclude that the theory is not \trit.

So Albert's analysis differs from the standard one on two accounts. First, there is the different notion of instantaneous state. In essence, Albert takes the instantaneous state at a time to be determined by the fundamental ontology (that is, by what there exists on the fundamental level according to the theory. For example, the ontology may be one of, say, particles or fields, so that the instantaneous state at a time consists of particle positions or field configurations at that time. On the other hand, on the standard account, the instantaneous state is such that it determines a unique solution to the equations of motion and as such may contain more variables than Albert's instantaneous state. For example, it may also include particle velocities, field velocities, \dots . Second, there is the notion of \trie, which is just the temporal order reversal of instantaneous states for Albert, whereas there might be an additional (involutive) state transformation at each instant according to the standard account. The examples of electrodynamics and quantum mechanics show that the second difference is essential for the different conclusion concerning the question of time reversal invariance. 

This issue is important because if a theory is not time reversal invariant, then it could be argued that time has an objective direction according to that theory (while it has no bearings on issues like the arrow of time \cite{albert00,wallace13}). There is a large body of interesting literature defending the standard time reversal transformations contra Albert and Callender \cite{earman02c,arntzenius04,malament04,north08,arntzenius09,roberts12,roberts17,roberts21}. In particular, efforts are made to make precise the notion of \trie. From the standard textbook account one may get the impression that the time reversal transformation of the instantaneous state is somewhat arbitrary and is chosen so as to make the theory \trit. It is a virtue of the Albert and Callender account that it does not depend on desiderata of this kind. Two other precise notions that stand out are that of `active time reversal' \cite{arntzenius04,arntzenius09,arntzenius11,arntzenius12} and a notion of Malament sometimes called `geometric time reversal' \cite{malament04,north08,arntzenius09}. An active transformation is defined through the notion of a passive transformation. Given geometric quantities that are expressed in a coordinate system with time $t$, a passive transformation expresses these quantities in a different coordinate system with time $t'=-t$. An active transformation keeps the coordinate system fixed but transforms the quantities as in the passive transformation. Geometric time reversal differs from active time reversal. In short, geometric time reversal corresponds to flipping just the temporal orientation, but keeping the geometrical objects fixed. Representations of these geometrical objects may depend on the temporal orientation and may hence change under geometric time reversal. As in the case of active time reversal, the way the instantaneous state transforms depends on the type of geometrical objects that exist. This makes that the transformation may be non-trivial so that time reversal may amount to more than the mere temporal order reversal of Albert and Callender. 

In this paper, we will not so much engage in the discussion on which is the better notion of time reversal invariance. Rather, the goal of the paper is to show that theories like electrodynamics and quantum mechanics can be considered time reversal invariant according to the Albert-Callender notion, provided a suitable ontology is chosen. Namely, the ontology of these theories is underdetermined (especially when it comes to field ontologies). Different ontologies yield different instantaneous states. So whether a theory counts as time reversal invariant depends on what is considered to be the ontology. Ontologies can be found such that the theories are time reversal invariant in the Albert-Callender sense.

The role of ontology has been emphasized before, notably in \cite{arntzenius09,allori15}. Arntzenius and Greaves show that different ontologies exist for which electrodynamics is time reversal invariant under geometric time reversal. They also consider Albert's view on electrodynamics. But while they consider it as internally coherent, they do not further pursue it because the electric and magnetic fields are regarded as a suitable ontology for a Newtonian space-time, but not for a relativistic one. Allori \cite{allori15} compares different views including those of Albert and Malament and argues that the difference lies (in part) in the choice of ontology. 

Since the Albert-Callender notion of time reversal invariance seems to involve mere reversal of temporal order, it seems stronger than the other notions of active and geometric time reversal. For theories that are standardly considered to be time-reversal invariant, ontologies may be found such that these theories are invariant under temporal order reversal. But not necessarily the other way around. For example, the standard model of particle physics is actually not time reversal invariant according to the standard notion. (It is merely CPT-invariant, that is, invariant under the joint transformation of charge conjugation, parity and time reversal.) But no choice of ontology will make the theory invariant under the temporal order reversal of Albert and Callender.

The ontologies we consider for electrodynamics and non-relativistic quantum mechanics also make the theories invariant under active and geometric time reversal. Unlike for electrodynamics, this has to our knowledge not yet been achieved for non-relativistic quantum mechanics. Roberts discusses the latter in detail and defends the usual transformation of the wave function under time reversal, but on different grounds \cite{roberts12,roberts17}. 

The outline of the paper is as follows. In the next section, we start with introducing the relevant notions. Then, in section \ref{ontology}, we will consider the ontological implications concerning time reversal invariance in the case of a scalar field. In section \ref{edqm}, we will consider ontologies for classical electrodynamics and quantum mechanics which make these theories \trit\ in the Albert-Callender sense. With these choices of ontology, the time reversal transformation happens to coincide with the standard one (just as it does in the case of Newtonian mechanics). There are also examples for which this is not the case. In section \ref{sed}, we will illustrate this with scalar electrodynamics, which describes a scalar field interacting with an electromagnetic field. An ontology will be presented for which the Albert and Callender notion of time reversal does not coincide with the standard one, but rather with the joint transformation of time reversal and charge conjugation. Finally, in section \ref{active}, a comparison is made with active and geometric time reversal in the case of electrodynamics. We conclude in section \ref{conclusion}.

\section{Instantaneous state and time reversal}
Let us first formalize some notions, following Albert \cite{albert00}. The instantaneous state at a certain time $t$ is denoted by $S(t)$. As mentioned before, for Albert, the instantaneous state at a time is determined by the ontology. For example, in the case of a particle ontology, the instantaneous state at a time is given by the particle positions at that time. (There might be non-dynamical variables such as for example the charges or masses of particles which may be part of the state, but which do not play a role in the arguments concerning time reversal invariance. Therefore, we will only explicitly include the dynamical variables in the state specification.) One may consider all kinds of other quantities such as velocities, accelerations, momenta, angular momenta, energies, etc., but these are not fundamental quantities; they can be derived from the trajectories of the particles. On the other hand, the standard notion of instantaneous state (in this context) includes extra variables at that time such that for a deterministic theory the instantaneous state together with the equations of motion determines a unique solution. For example, in the case of a particle ontology these variables could be the velocities or the momenta. We will add the subscripts $a$ and $s$ and write $S_a(t)$ and $S_s(t)$ to refer to respectively Albert's notion and the standard notion of state. (Albert elaborates more on the notion, but this suffices for our purposes.) 

For a given history $t \to S(t)$, the time-reversed history is denoted by $t \to T(S)(t)$. For Albert, the time-reversed history is $t \to T_a(S_a)(t)$, with $T_a(S_a)(t)= S_a(-t)$.{\footnote{We consider only time-translation invariant theories so that there is nothing special about $t=0$ in the definition of a time-reversed history.}} So time reversal is merely a reversal of the temporal order of the instantaneous states. It also induces a transformation of the non-fun\-da\-mental quantities like velocities and momenta, etc., which may amount to more than their mere order reversal. According to the standard notion, given a history $t \to S_s(t)$, the time-reversed history is $T_s(S_s)(t)=S^T_s(-t)$, where the superscript $T$ denotes some additional involutive operation on each instantaneous state in addition to the order reversal.

A theory is time reversal invariant if for each dynamically allowed history, that is, each possible solution to the equations of motion, its time-reversed history is also dynamically allowed.

Let us give some examples. First consider Newtonian mechanics, where the ontology is given by point-particles with positions $({\bf X}_1,\dots,{\bf X}_n)$. The equations of motion read
\be
m_k \frac{d^2{\bf X}_k}{dt^2} = -{\boldsymbol \nabla}_k V({\bf X}_1,\dots,{\bf X}_n),
\en
with $m_k$ the mass of the $k$-th particle and $V$ a potential which depends on just the positions (not on time or the velocities). According to the standard notion, the instantaneous state at a time $t$ is the collection of positions and velocities ${\bf V}_k$ at that time, where
\be
{\bf V}_k(t) = \frac{d {\bf X}_k(t)}{dt},
\label{vel}
\en
so that
\be
S_s (t) =  ({\bf X}_1(t),\dots,{\bf X}_n(t),{\bf V}_1(t),\dots,{\bf V}_n(t)).
\en
The time reversal operation is
\be
T_s: S_s(t) \to   S^T_s (-t) =  ({\bf X}_1(-t),\dots,{\bf X}_n(-t),-{\bf V}_1(-t),\dots,-{\bf V}_n(-t)).
\label{state}
\en
Since this operation maps solutions to solutions, Newtonian mechanics is time reversal invariant according to the standard notion.{\footnote{For this conclusion to obtain, it is important that the potential does not depend on time or the particle velocities.}} According to Albert, the instantaneous state at time $t$ is
\be
S_a (t) =  ({\bf X}_1(t),\dots,{\bf X}_n(t)).
\en
It does not include the instantaneous velocities, but these are determined by the collection of instantaneous states, that is, as the rates of change of the positions, by \eqref{vel}. The time reversal transformation is
\be
T_a: S_a(t) \to   S_a (-t) =  ({\bf X}_1(-t),\dots,{\bf X}_n(-t)),
\en
with the induced transformation of the velocities as in \eqref{state}. The upshot is that also according to Albert, Newtonian mechanics is time reversal invariant.

Let us now turn to classical electrodynamics. Consider an ontology given by point-particles together with an electric and magnetic field ${\bf E}({\bf x},t)$ and ${\bf B}({\bf x},t)$.{\footnote{The electric and magnetic field can also be written in terms of the electromagnetic tensor
\be
F^{\mu \nu} = 
\begin{pmatrix} 
0 & E_1 & E_2 & E_3 \\
-E_1 & 0 & B_3 & -B_2 \\
-E_2 & -B_3 & 0 & B_1 \\
-E_3 & B_2 & -B_1 & 0 
\end{pmatrix}.
\label{fmunu}
\en
}} 
The laws of motion are given by the Lorentz force law{\footnote{We assume units such that $c = \hbar = 1$ throughout.}}
\be
\frac{d}{dt} \left( m_{r,k} \frac{ d{\bf X}_k}{dt}\right) =  e_k \left[{\bf E}({\bf X}_k,t ) + \frac{d{\bf X}_k}{dt} \times  {\bf B}({\bf X}_k,t )  \right]    ,
\en
where $m_{r,k}$ is the relativistic mass of the $k$-th particle and $e_k$ its charge, together with Maxwell's equations
\be
{\boldsymbol \nabla} \cdot {\bf E} = \rho, \qquad {\boldsymbol \nabla} \cdot {\bf B} = 0,
\en   
\be
{\boldsymbol \nabla} \times {\bf E} = - \frac{\pa {\bf B}}{\pa t}, \qquad{\boldsymbol \nabla}\times {\bf B} =  {\bf J} + \frac{\pa {\bf E}}{\pa t},
\en
where $\rho({\bf x},t) = \sum_k e_k \delta({\bf x} - {\bf X}_k(t))$ and ${\bf J}({\bf x},t) = \sum_k e_k \frac{d{\bf X}_k(t)}{dt}\delta({\bf x} - {\bf X}_k(t))$ are respectively the charge density and the charge current. According to the standard account, the instantaneous state is
\be
S_s (t) =  ({\bf X}_1(t),\dots,{\bf X}_n(t),{\bf V}_1(t),\dots,{\bf V}_n(t), {\bf E}({\bf x},t), {\bf B}({\bf x},t))
\en
and under time reversal
\be
T_s:S_s(t) \to   S^T_s (-t) =   ({\bf X}_1(-t),\dots,{\bf X}_n(-t),-{\bf V}_1(-t),\dots,-{\bf V}_n(-t), {\bf E}({\bf x},-t), -{\bf B}({\bf x},-t)).
\label{usual}
\en
It is crucial that the magnetic field flips sign under this operation as it guarantees the invariance of the equations of motion.

Albert takes the instantaneous state to be
\be
S_a (t) =  ({\bf X}_1(t),\dots,{\bf X}_n(t), {\bf E}({\bf x},t), {\bf B}({\bf x},t))
\en
and under time reversal
\be
T_a:S_a(t) \to   S_a (-t) =   ({\bf X}_1(-t),\dots,{\bf X}_n(-t),{\bf E}({\bf x},-t), {\bf B}({\bf x},-t)).
\label{30}
\en
There is no sign flip of the magnetic field; it is not the rate of change of anything. As such, Albert concludes that the equations of motion are not time reversal invariant; the transformation \eqref{30} is not a symmetry of the equations of motion (it does not map solutions to solutions). The standard time reversal transformation \eqref{usual} is still a symmetry of the equations of motion, but for Albert this does not amount to time reversal symmetry.

In non-relativistic quantum mechanics the situation is similar and is detailed by Callender \cite{callender00}. To avoid the interpretational issues that arise in this context, we will regard the Schr\"odinger equation as just a classical field equation. For simplicity, we will also consider just a single ``particle''. The field ontology is then represented by the wave function $\psi({\bf x},t)$ and the Schr\"odinger equation is
\be
\ii \frac{\partial \psi({\bf x},t)}{\partial t} = -\frac{1}{2m} \nabla^2\psi({\bf x},t) + V({\bf x})\psi({\bf x},t).
\label{schr}
\en
The instantaneous state is
\be
S_s (t) = S_a (t) = \psi({\bf x},t).
\en 
The Schr\"odinger equation is invariant under the standard time reversal operation
\be
T_s:S_s (t) \to  S^T_s (-t) = \psi^*({\bf x},-t),
\label{31}
\en
but not under the mere reversal of temporal order of instantaneous states
\be
T_a:S_a (t) \to  S_a (-t) = \psi({\bf x},-t).
\en

\section{Ontological underdetermination}\label{ontology}
There tends to be an underdetermination in the ontology of physical theories. Newtonian mechanics is usually regarded as a theory about point-particles. But one could consider different possible ontologies. In particular, there are ontologies that make that the theory is no longer invariant under mere temporal order reversal. For example, take the ontology to be given by point-particles, with positions ${\bf X}_k$, endowed with vectors ${\bf P}_k$ at the particle locations, and take the dynamics to be given by
\be
m_k \frac{d {\bf X}_k }{dt} = {\bf P}_k, \qquad \frac{d {\bf P}_k}{dt} = -{\boldsymbol \nabla}_k V({\bf X}_1,\dots,{\bf X}_n).
\label{fp}
\en
This is of course recognized as the phase space formulation of Newtonian mechanics. But while the ontology is (usually) still taken to consist of just the point-particles, with \eqref{fp} corresponding to a particular way of expressing the particle dynamics, we consider here an alternative ontology given by the point-particles together with the vectors ${\bf P}_k$. That the ${\bf P}_k$ are related to the rates of change of the positions is not taken as a kinematical fact, but as a dynamical fact.{\footnote{To appreciate the difference, consider a different (physically unmotivated) dynamics for which this relation no longer holds, like $d {\bf X}_k /dt= 0$, $d {\bf P}_k /dt= 0$.}} With this ontology, the instantaneous state is $S_a(t) = ({\bf X}_1(t),\dots,{\bf X}_n(t),{\bf P}_1(t),\dots,{\bf P}_n(t))$ and the theory is no longer invariant under mere temporal order reversal. 

Consider now the theory of a real scalar field $\phi({\bf x},t)$, satisfying the Klein-Gordon equation
\be
\pa_\mu \pa^\mu \phi + m^2 \phi=0.
\label{o1}
\en
The ontology is given by the scalar field $\phi({\bf x},t)$. But one could also consider a phase space representation, with an ontology given by the fields $(\phi({\bf x},t),\pi({\bf x},t))$, satisfying
\be
\frac{\pa \phi}{\pa t} = \pi, \qquad  \frac{\pa \pi}{\pa t} =  \nabla^2 \phi - m^2 \phi.
\en
As with the alternative ontology for Newtonian mechanics above, the fields $\phi({\bf x},t)$ and $\pi({\bf x},t)$ should be taken as ontologically independent fields. That the field $\pi$ happens to agree with the velocity $\pa \phi/\pa t$ is merely a dynamical fact and not a kinematical one. 

The scalar field theory can also be written in terms of a 5-component spinor $\psi({\bf x},t)$ which satisfies the Kemmer equation \cite{kemmer39,akhiezer65}
\be
\ii  \beta^\mu \pa_\mu\psi - m\psi =0.
\label{o3}
\en
This is a Dirac-like equation, which is manifestly Lorentz invariant, just as the Klein-Gordon equation \eqref{o1}. This theory is completely equivalent to the Klein-Gordon theory. (In a particular representation of the Kemmer matrices $\beta^\mu$, the Kemmer equation implies $\psi = (\pa_\mu \phi, m \phi)^T$.) Despite the equivalence, this form of the theory is hardly used, due to its greater complexity. However, this is not a reason not to consider an ontology in terms of the Kemmer spinor. Actually, when it comes to spin-1/2 particles, the Dirac equation which is a first-order differential equation (which is the analogue of \eqref{o3}) is the one commonly used, instead of the somewhat simpler second-order Van der Waerden equation for a two-component spinor \cite{sakurai67} (which is the analogue of \eqref{o1}).

So for the scalar field theory, we have considered three possible candidates for the ontology and hence for the instantaneous state $S_a$. Namely,
\be
S^{(1)}_a(t) = \phi({\bf x},t) ,\qquad  S^{(2)}_a(t) = (\phi({\bf x},t),\pi({\bf x},t)), \qquad S^{(3)}_a(t) = \psi({\bf x},t).
\en
Only the first one yields time reversal invariance under temporal order reversal. The time reversal operation $T_a$ in this case also corresponds to the standard one. According to the standard notion, the theory is time reversal invariant for all these choices of ontology. (Similarly, in the case of a (free) spin-1/2 particle, for which the state $S_a$ can be taken to be a Dirac spinor or a Van der Waerden spinor, only the latter will amount to invariance under temporal order reversal.)  

So whether a theory is invariant under temporal order reversal depends on the choice of ontology. In the case of a classical field theory, different possible ontologies seem possible with no clear physical preference. (Even the requirement of manifest Lorentz invariance leaves options $S^{(1)}_a$ and $S^{(3)}_a$.) In the next section, we will show that the underdetermination of the ontology in the case of classical electrodynamics and non-relativistic quantum mechanics can be exploited to choose one such that the theory is invariant under mere temporal order reversal.

\section{Ontology of electrodynamics and quantum mechanics}\label{edqm}
Maxwell's equations imply{\footnote{The action of $1/\nabla^2$ is defined in terms of the Green function of the Laplacian as $(1/\nabla^2) f({\bf x}) = -\frac{1}{4\pi}\int d^3 y f({\bf y})/|{\bf x}-{\bf y}|$. The expression \eqref{mag} follows if the fields fall off sufficiently fast at spatial infinity.}}
\be
{\bf B} = - \frac{1}{\nabla^2} {\boldsymbol \nabla} \times \left(  {\bf J} + \frac{\pa {\bf E}}{\pa t} \right).
\label{mag}
\en
This expression can be used to eliminate the magnetic field from Maxwell's equations and the Lorentz force law. Maxwell's equations are then expressed as
\be
{\boldsymbol \nabla} \cdot {\bf E} = \rho, \qquad \frac{\pa^2 {\bf E}}{\pa t^2} - \nabla^2  {\bf E} = - {\boldsymbol \nabla} \rho - \frac{\pa {\bf J}}{\pa t} .
\label{second}
\en
The resulting formulation of the theory is completely equivalent to the original one, where the latter can be obtained using \eqref{mag} as a definition of the magnetic field. The number of equations of motion is halved, but the equations have become second order in the time derivatives rather than first order. The reformulation suggests that the ontology of the electromagnetic field can be taken to be just the electric field, so that 
\be
S_a (t) =  ({\bf X}_1(t),\dots,{\bf X}_n(t), {\bf E}({\bf x},t)),
\en
with the field equations given by \eqref{second}. This theory is invariant under mere temporal order reversal. There is no problem with the magnetic field because it is simply not part of the ontology (and the theory). The magnetic field could be defined as in \eqref{mag}, in terms of the particles and the electric field. From that definition it follows that under temporal order reversal of the state $S_a$, the magnetic field will flip sign. So on this view the magnetic field does play the role of a velocity, since it is a linear combination of the rates of change of the particle positions (through the charge current) and the electric field.

Note that we could also have eliminated the electric field in terms of the magnetic field. But then the resulting theory would not be invariant under temporal order reversal. So there is no technical reason why it is more natural to assume the ontology to be given by the electric field rather than the magnetic field.

Nevertheless, there is an issue with this ontology. It is an ontology that is suitable for Newtonian space-time, but not so much for a Minkowski space-time, which is the natural space-time in this context due to the Lorentz invariance of electrodynamics (see also \cite{arntzenius09}). So rather than having a 3-vector as constituting the fundamental ontology, it would be more desirable to have Lorentz-covariant objects, like the electromagnetic tensor \eqref{fmunu}. However, such an ontology does not make the theory invariant under mere temporal order reversal. The same holds for the ontology considered by Malament \cite{malament04}, where the electromagnetic field is a map from tangent lines to forces. We will discuss this further in section \ref{active} when comparing to other notions of time reversal.

A manifestly Lorentz invariant theory that is invariant under temporal order reversal could be obtained by completely removing the fields from the ontology, so that only the particles remain. This is attempted in the Wheeler-Feynman theory \cite{spohn04,lazarovici18}. To see how this theory is obtained, consider the covariant form of electrodynamics
\be
\pa_\mu F^{\mu \nu}(x) = \sum_k j^\nu_k(x) , \qquad m_k \frac{d^2 X^\mu_k(s_k)}{ds^2_k} = e_k F^{\mu}{}_\nu(X_k(s_k)) \frac{d X^\nu_k(s_k)}{ds_k},
\label{99}
\en
where $X^{\mu}_k(s_k)$ is the world line of the $k$-th particle, parameterized by its proper time $s_k$, $A^\mu(x)$ is the electromagnetic potential and $F^{\mu \nu}=\pa^\mu A^\nu -\pa^\nu A^\mu $ is the electromagnetic tensor, with $E_i = F_{0i}$ and $B_i= - \epsilon_{ijk}F^{jk}/2$ (see \eqref{fmunu}), and $j^\mu_k(x) =  e_k \int ds \frac{d X^\mu_k(s)}{ds} \delta(x - X_k(s))$ the charge current produced by the $k$-th charge. Assuming the Lorenz gauge $\pa_\mu A^\mu = 0$, the Maxwell equations can be written as
\be
\square A^\mu (x) =  \sum_k j^\mu_k(x) .
\label{99.1}
\en
The potential can be decomposed as $A^\mu_F + A^\mu_M$ with $A^\mu_F$ a field satisfying the free Maxwell equations $\square A^\mu_F (x) = 0$ and
\be
A^\mu_M(x) = \frac{1}{\square} \sum_k j^\mu_k(x),
\label{100}
\en
where $1/\square$ denotes a convolution with a Green's function $G$ of the d'Alembertian. There are various choices for $G$; one could take the retarded Green's function $G_-$, the advanced one $G_+$ or linear combinations.{\footnote{More explicitly, the Green's function $G$ is a solution to $\square G({\bf x},t;{\bf x}',t') = \delta({\bf x} - {\bf x}') \delta(t-t')$ and \eqref{100} amounts to $A^\mu_M(x) = \int d^4 x' G(x;x')j^\mu(x')$. The retarded and advanced solutions read $G_\pm({\bf x},t;{\bf x}',t') = -\frac{1}{4\pi} \delta(t-t' \mp |{\bf x} - {\bf x}'|) \theta(t-t')$, with $\theta$ the Heavyside step function.}} Different choices imply different free fields $A^\mu_F$. Wheeler and Feynman chose $G = (G_+ + G_-)/2$. Eq.\ \eqref{100} is then taken as a definition of $A^\mu_M$, rather than as (part of) a dynamical equation. In the Lorentz force law in \eqref{99}, the self-force is subtracted to avoid infinities (so that the sum ranges over $l \neq k$ in \eqref{99.2}). Furthermore, it is assumed there are no free fields, so that $A^\mu_F=0$. In this way, the fields are completely eliminated from the theory. There are just particles, satisfying the equations of motion
\be
m_k \frac{d^2 X^\mu_k(s_k)}{ds^2_k} = e_k \sum_{l \neq k}\left[ \pa^\mu \frac{1}{\square} j_{l\, \nu}(X_k(s_k)) - \pa_\nu \frac{1}{\square}    j^\mu_l(X_k(s_k))         \right] \frac{d X^\nu_k(s_k)}{ds_k}.
\label{99.2}
\en
The instantaneous state is now
\be
S_a (t) =  ({\bf X}_1(t),\dots,{\bf X}_n(t))
\en
and it can easily be checked that the theory is invariant under temporal order reversal.{\footnote{To establish this time reversal invariance, it is crucial that the Green's function $G = (G_+ + G_-)/2$ is chosen in \eqref{100}, since this function satisfies $G({\bf x},t;{\bf x}',t') = G({\bf x},-t;{\bf x}',-t')$. For any other choice of $G$, the dynamics \eqref{99.2} will not be time reversal invariant. In addition, the transformation law of $A_M$ \eqref{101} only holds for the Wheeler-Feynman choice. It is also crucial for the time reversal invariance that the free field is assumed zero. If it is not zero and assumed to be part of the ontology, then while the free Maxwell equations $\square A^\mu_F =0$ are invariant under temporal order reversal, the Lorentz force law will not be. }} 

There are no fields, but if one defines $A^\mu$ through \eqref{100} (with the Wheeler-Feynman choice of Green's function), the time reversal transformation $T_a$ on the state $S_a$ induces the follow transformation of $A^\mu_M$
\be
A^0_M({\bf x},t) \to A^0_M({\bf x},-t) , \qquad A^i_M({\bf x},t) \to - A^i_M({\bf x},-t).
\label{101}
\en
This transformation agrees with the standard transformation of the vector potential found in textbooks. It further induces the standard transformations of the electric and magnetic field if one defines them through the usual definitions, including a sign flip of the magnetic field.{\footnote{Interestingly, in his defense of the standard notion of time reversal invariance, Earman also considers $A^\mu_M$ \cite{earman02c}. He proposes to take \eqref{99.1} as `the definition of the four-potential arising from [the current]'. This could be read in the sense considered here, namely that there is no independent reality for the field. However, Earman seems to have had merely the intention of showing that if one accepts the usual transformation of the particles, then one should also accept the usual transformation of the vector potential. But this is akin to stating that the vector potential should transform the usual way, just to make the theory time reversal invariant. Because if $A^\mu_M$ (or the electromagnetic field) is taken as part of the ontology, then \eqref{99.1} should be taken as a law.}}  

There is debate about the empirical adequacy of the theory, but the important point for our purposes is that it is a theory that is manisfestly Lorentz invariant and that it is invariant under temporal order reversal. Other choices of ontologies may be possible that achieve this and perhaps also include a free field.

Allori \cite{allori15} considers yet another option to have invariance under temporal order reversal, which she attributes to Horwich \cite{horwich87}. On this view, the electric and magnetic field are not part of the fundamental ontology. There is just a particle ontology, like in the Wheeler-Feynman theory. But unlike in the latter, the electromagnetic field still appears in Horwich's account of the theory. But the field has a nomological character rather than an ontological one, that is, the field merely plays a role in the dynamics of the particles. The particles are said to constitute the `primitive ontology'. The fields then transform the way they do under time reversal just to have the primitive ontology transform the right way. The approach we consider here is different. We have not relegated some parts of the standard ontology to the nomological domain, but rather we have eliminated them completely (in particular also from the dynamics). For example, in our first proposal, the magnetic field was no longer part of the theory, neither on the ontological nor on the nomological level.

The non-relativistic Schr\"odinger equation can be dealt with similarly. Writing $\psi=\psi_r + \ii \psi_i$, with $\psi_r$ and $\psi_i$ real, the Schr\"odinger equation \eqref{schr} amounts to the following set of coupled differential equations
\be
\pa_t \psi_r = -H \psi_i,
\label{re}
\en
\be
\pa_t \psi_i = H \psi_r,
\label{im}
\en
where $H = -\frac{1}{2m} \nabla^2 +V({\bf x})$ is the Hamiltonian operator. Taking the time derivative of \eqref{re} and using \eqref{im}, leads to{\footnote{This equation has been considered by a number of people, including Schr\"odinger himself \cite{callender20}. This equation is also encountered in the reduced phase space formulation of the Schr\"odinger theory~\cite{struyve10}. In the reduced phase space formulation, $\psi_r$ and $\psi_i$ become canonically conjugate variables. An inverse Legendre transformation leads to the Lagrangian for just $\psi_r$, whose Euler-Lagrange equation corresponds to \eqref{second2}. Actually, the second-order equations \eqref{second} for electromagnetism could be obtained similarly since the electric and magnetic field are (approximately) canonically conjugate. The analogy between the second-order equations for electromagnetism and non-relativistic quantum mechanics also shows up if the Riemann-Silberstein vector ${\bf F} = {\bf E} + \ii {\bf B}$ is used for the electromagnetic field \cite{bialynicki-birula96}. The free Maxwell equations imply the Schr\"odinger-like equation $\ii \pa {\bf F}/ \pa t = {\boldsymbol \nabla} \times {\bf F}$, together with the constraint ${\boldsymbol \nabla} \cdot {\bf F} =0$. Eliminating the complex part of ${\bf F}$ then corresponds to eliminating the magnetic field.}}
\be
\frac{\pa^2 \psi_r }{\pa t^2} =  - H^2 \psi_r.
\label{second2}
\en
In this way the imaginary part is eliminated. Rather than taking the ontology to be given by $\psi$, it can be taken to be just $\psi_r$, satisfying the real wave equation \eqref{second2}. The theory is still equivalent to the Schr\"odinger equation, by defining {\footnote{The inverse of the Hamiltonian operator $1/H$ can be defined in terms of the Green's function for $H$. We have hereby assumed that the inverse of $H$ does indeed exist. It might not exist if the spectrum includes zero. However, if the spectrum is bounded from below, this can easily be taken care of by shifting the spectrum, via the Hamiltonian $H'=H + E$, with $E$ a constant such that the spectrum of $H'$ no longer includes zero. On the level of the Schr\"odinger equation this shift leads to an equivalent theory, since it merely entails a phase shift of the solutions, given by $\psi'=\ee^{-\ii Et}\psi$.}}
\be
\psi_i = \frac{1}{H} \pa_t \psi_r .
\label{psii}
\en

The equation \eqref{second2} is second order in the time derivative, so that the theory is time reversal invariant under mere temporal order reversal.{\footnote{Quantum mechanics actually entails much more than just the Schr\"odinger equation. What this is, depends on the version of quantum mechanics. Let us briefly say something about the time reversal invariance for the three main attempts that solve the measurement problem, namely the many worlds theory, spontaneous collapse theories and Bohmian mechanics. In the many worlds theory, the ontology is given by just the wave function and hence it can be considered invariant under temporal order reversal by taking the ontology to be given by just the real part of the wave function. In spontaneous collapse theories, the Schr\"odinger evolution of the wave function is interrupted by collapses which are stochastic. This entails further discussion of the notion of time reversal invariance which we will not consider. Usually, the theory is not considered time reversal invariant even in the standard sense \cite{callender00}, but see also \cite{bedingham17}. In Bohmian mechanics there are also actual point-particles in addition to the wave function. The Bohmian dynamics is time reversal invariant in the standard sense \cite{duerr92a2}, with $X(t) \to X(-t)$ and $\psi(x,t) \to \psi^*(x,-t)$, and under temporal order reversal if the above ontology for the wave function is adopted.}}  The usual time reversal transformation \eqref{31} for $\psi$ is recovered since the definition \eqref{psii} for the imaginary part $\psi_i$ entails that it transforms as the time derivative of $\psi_r$. So $\psi_i$ roughly plays the role of a field velocity.

\section{Time reversal invariance with different time reversal transformations}\label{sed}
In the previous section, examples of ontologies were provided that make electrodynamics and the non-relativistic Schr\"odinger equation invariant under mere temporal order reversal. This was done by respectively removing the magnetic field and the imaginary part of the wave function from the ontology. In these cases, the transformation was in agreement with the standard time reversal transformation. That is, the temporal order reversal of the fundamental variables was just the standard time reversal transformation, as was the induced transformation of the non-fundamental variables. However, this need not always be the case. The transformations may disagree, yet yield time reversal invariance under both notions. Consider for example a complex scalar field $\phi$ satisfying the Klein-Gordon equation \eqref{o1}, which describes a charged spinless field. According to the standard notion of time reversal, the field should transform as $\phi({\bf x},t) \to \phi^*({\bf x},-t)$, whereas under mere temporal order reversal, taking $S_a(t)= \phi({\bf x},t)$, $T_a: \phi({\bf x},t) \to \phi({\bf x},-t)$. So there is disagreement about what counts as time reversal. Yet, both transformations are symmetries of the Klein-Gordon equation (they map solutions to solutions) and hence according to both notions the theory is time reversal invariant. (The same is true for the Van der Waerden equation that was mentioned in section \ref{ontology}. With the Van der Waerden spinor as ontology, the theory is invariant under temporal order reversal, even though this does not amount to the standard time reversal operation.)

The previous example can be extended to include an electromagnetic field. In terms of the scalar field $\phi$ and the vector potential $A^\mu$, this theory (scalar electrodynamics) has the equations of motion
\be
D_\mu D^\mu \phi + m^2 \phi =0 ,\qquad \pa_\mu F^{\mu \nu} = j^\nu,
\en 
where $D_\mu = \pa_\mu + \ii e A_\mu$ is the covariant derivative and 
\be
j^\mu = \ii e \left[ \phi^* D^\mu \phi - \phi (D^\mu \phi)^* \right]
\en
is the charge current. The theory is invariant under the standard time reversal operation
\be
\psi({\bf x},t) \to \psi^*({\bf x},-t), \qquad A^0({\bf x},t) \to A^0({\bf x},-t) , \qquad A^i({\bf x},t) \to - A^i({\bf x},-t). 
\label{200}
\en
But taking $\phi$ and $A^\mu$ as the ontology does not make this theory invariant under temporal order reversal. 

Consider now the temporal gauge $A_0=0$. Then the equations of motion are
\be
\frac{\pa^2 \phi}{\pa t^2} - {\bf D}\cdot {\bf D} \phi +  m^2 \phi =0 , \quad \square {\bf A} + {\boldsymbol \nabla } ({\boldsymbol \nabla } \cdot  {\bf A}) = {\bf j}, \quad - {\boldsymbol \nabla } \cdot  \frac{\pa {\bf A}}{\pa t} = j_0,
\en
with now ${\bf D} = {\boldsymbol \nabla} - \ii e {\bf A}$ and the charge density and 3-current respectively given by 
\be
j_0 = \ii e \left( \phi^* \frac{\pa \phi}{\pa t} - \phi  \frac{\pa \phi^*}{\pa t} \right), \qquad {\bf j} = \ii e \left[ \phi {\bf D} \phi^* - \phi^* {\bf D} \phi \right].
\label{205}
\en
Taking the state to be
\be
S_a(t) = \left( \phi({\bf x},t),{\bf A}({\bf x},t) \right),
\en
then it is readily checked that the theory is invariant under temporal order reversal $T_a: S_a(t) \to S_a(-t)$. Nevertheless, this is a symmetry different from the standard time reversal symmetry \eqref{200}. The transformation $T_a$ considered here actually corresponds to the joint time reversal (T) and charge conjugation (C) in the standard picture. Namely, under charge conjugation, one has $\phi \to \phi^*$ and $A^\mu \to - A^\mu$. So in this case, the temporal order reversal transformation $T_a$ amounts to the joint $TC$ transformation of the standard picture. (This explains why under $T_a$, the charge current in \eqref{205} flips sign. That is, it transforms as $j_0({\bf x},t) \to -j_0({\bf x},-t)$, ${\bf j}({\bf x},t) \to -{\bf j}({\bf x},-t)$.) A similar point is made in \cite{arntzenius09,arntzenius11,arntzenius12}, where it is argued that the joint CPT transformation is really a PT transformation.

\section{Active and geometric time reversal}\label{active}
So far we have been dealing with the Albert and Callender notion of time reversal which amounts to the mere reversal of the temporal order of the instantaneous states. It is worth comparing this notion with those of active and geometric time reversal that were mentioned in the introduction. The difference comes about especially in the case of an ontology with tensorial objects in space-time. Such objects are the natural ontological objects to consider in a Lorentz invariant theory like electrodynamics. However, under mere reversal of temporal order, they do not transform in a way to make electrodynamics time reversal invariant. (This is why we have achieved time-reversal invariance in section \ref{edqm} only by considering an ontology in terms of the electric field as a spatial 3-vector or by considering the Wheeler-Feynman ontology in terms of just particles.) To see this, consider first a 4-vector field $V^\mu(x)$. Under mere temporal order reversal, the vector field transforms as
\be
V^\mu({\bf x},t)  \to V^\mu({\bf x},-t).
\label{300}
\en 
So the instantaneous state at a time $t$ is taken to be the vector field at that time and the temporal order of these states is reversed. As a concrete example, consider the vector potential $A^\mu$. Transforming $A^\mu$ as in \eqref{300} does not amount to a symmetry of electrodynamics (and neither does it imply the transformation \eqref{30} for the electric field since it will flip sign). The same conclusion holds for the electromagnetic tensor $F^{\mu \nu}$. Under mere temporal order reversal, $F^{\mu \nu}({\bf x},t)  \to F^{\mu \nu}({\bf x},-t)$, and as such the electric and magnetic field $E_i = F_{0i}$ and $B_i= - \epsilon_{ijk}F^{jk}/2$ transform as in \eqref{30}, without the sign flip for the magnetic field. But as discussed before this is not a symmetry.

Let us now turn to active time reversal \cite{arntzenius04,arntzenius09,arntzenius11,arntzenius12}. For this we first need to consider the passive transformation which is obtained by taking a different coordinate system with $t' = -t$. In terms of the new coordinates, the vector field $V^\mu(x)$ reads
\be
V'^\mu ({\bf x},t') =  (-V^0({\bf x},-t), V^i({\bf x},-t)).
\en
The active transformation then keeps the coordinate system fixed but changes the vector field $V^\mu(x)$ to a different one with the same form as in the passive transformation: 
\be
V^\mu({\bf x},t)  \to (-V^0({\bf x},-t), V^i({\bf x},-t)).
\en
Applying this to the vector potential, this transformation induces the following transformation of the electric and magnetic field \cite{arntzenius04}:
\be
{\bf E}({\bf x},t) \to - {\bf E}({\bf x},-t)  , \qquad {\bf B}({\bf x},t) \to{\bf B}({\bf x},-t). 
\label{305}
\en
This is not the standard transformation given in \eqref{usual}. Actually, it corresponds to the joint transformation of standard time reversal and charge conjugation (as also encountered in the previous section). The same holds for the electromagnetic tensor $F^{\mu \nu}$. Under active time reversal, 
\be
F^{00} ({\bf x},t)  \to F^{00} ({\bf x},-t), \qquad F^{0i} ({\bf x},t)  \to -F^{0i} ({\bf x},-t), \qquad F^{ij} ({\bf x},t)  \to F^{ij} ({\bf x},-t)
\en
and hence again \eqref{305} follows. Electrodynamics with point charges is then invariant under active time reversal if the world lines also have an intrinsic direction which flips under time reversal, in accordance with Feynman's view of anti-particles \cite{arntzenius09}. If instead of point charges, matter consists of a complex scalar field $\phi({\bf x},t)$, then under active time reversal $\phi({\bf x},t) \to \phi({\bf x},-t)$, which again amounts to the joint transformation of the standard time reversal and charge conjugation transformations, given respectively by $\phi({\bf x},t) \to \phi^*({\bf x},-t)$ and $\phi({\bf x},t) \to \phi^*({\bf x},t)$.  These observations were used to argue that the CPT theorem is actually a PT theorem, removing the mystery why charge conjugation should have anything to do with space-time symmetries \cite{arntzenius11,arntzenius12}.

Malament's geometric time reversal \cite{malament04} employs the notion of a temporal orientation, which is represented by a time-like vector field $\tau^\mu$. Geometrical objects are not changed under time reversal, unlike in the case of temporal order reversal and active time reversal, but merely the temporal orientation is flipped: $\tau^\mu \to -\tau^\mu$. Representations of the geometrical objects may depend on the temporal orientation and may transform non-trivially as a result. Take for example the charge current. Malament considers the fundamental geometrical object to be $J$ which is a linear map from tangent lines to scalars, representing the charge densities. Using the temporal orientation, this map can be represented by a 4-vector field $J^\mu$. Namely, a (time-like) tangent line determines two unit tangent vectors $\xi^\mu$ and $-\xi^\mu$, where $\xi^\mu$ is future-directed relative to $\tau^\mu$, that is, $\xi^\mu \tau_\mu >0$, and $-\xi^\mu$ is past-directed relative to $\tau^\mu$. The 4-vector field $J^\mu$ is then defined as the map from the future-directed (relative to $\tau$) tangent vectors to scalars, given by $\xi^\mu \to J^\mu \xi_\mu$. Under geometric time reversal, the geometrical object $J$ is held fixed, while $\tau^\mu \to -\tau^\mu$. As a result, the vector field representation $J^\mu$ changes as $J^\mu \to -J^\mu$. On Malament's view the electromagnetic tensor is again a representation of some more fundamental object, namely a map $F$ from tangent lines to forces. Together with a temporal orientation this map $F$ determines a tensor $F^{\mu \nu}$, which is understood as a map from future-directed tangent vectors to forces. Under geometric time reversal, $F^{\mu \nu} \to - F^{\mu \nu}$. Together with the transformation of the charge current, this implies that Maxwell's equations are invariant under geometric time reversal. What does geometric time reversal entail for the transformations of the electric and magnetic field? As Malament explains, to define the electric and magnetic field, one needs to introduce a volume element $\epsilon_{\mu \nu \al \beta}$ and a `frame' which is represented by a constant vector field $\eta^\mu$, and which determines a space-time splitting, with the surfaces normal to $\eta^\mu$ corresponding to the spatial slices. The electric and magnetic field are then defined as $E^\mu = F^{\mu \nu} \eta_\nu$ and $B^\mu = \epsilon^{\mu \nu \al \beta} \eta_\nu F_{ \al \beta}$ (which are vectors tangential to the spatial hyperplanes). Malament argues that time reversal flips the sign of both $\eta^\mu$ and $\epsilon_{\mu \nu \al \beta}$, so that $E^\mu  \to E^\mu $ and $B^\mu  \to - B^\mu $, which amounts to the standard transformations of the electric and magnetic fields.

Arntzenius and Greaves offer also another possible ontology for which electrodynamics is invariant under geometric time reversal \cite{arntzenius09}. It is an ontology suggested by the Feynman picture where anti-particles are regarded as particles moving backwards in time. Arntzenius and Greaves take the ontology to be given by the electromagnetic tensor $F^{\mu \nu}$ which, unlike Malament's choice, is now regarded as fundamentally a map from four-vectors to four-vectors. So the tensor does not depend on the temporal orientation and hence is invariant under geometric time reversal. The other difference with Malament is that the world lines of the particles are assumed to carry a direction, which may or may not align with the temporal orientation, but which in any case is independent of it and hence also does not change under geometric time reversal.

The theories considered in section \ref{edqm} that were invariant under temporal order reversal are also invariant under active and geometric time reversal. Consider for example, electrodynamics with only the electric field and the point-particles as the ontology. The electric field is taken to be a spatial 3-vector rather than as derived from a vector potential or electromagnetic tensor and hence implies the active time reversal transformation ${\bf E}({\bf x},t) \to {\bf E}({\bf x},-t)$. This transformation agrees with the temporal order reversal. Likewise for the positions of the particles. So the theory is invariant under active time reversal. Under geometric time reversal the electric field and the positions are left invariant, they do not depend on the temporal orientation, but the frame $\eta^\mu$ changes to $-\eta^\mu$, so that the time derivative in the equations of motion (which depends on the frame) changes as $\pa/\pa t \to - \pa/\pa t$. This makes the theory invariant under geometric time reversal. Also the Wheeler-Feynman theory with its particles-only ontology and non-relativistic quantum mechanics with $\psi_r$ as the ontology are both invariant under active and geometric time reversal.

\section{Conclusion}\label{conclusion}
Prima facie, according to the Albert-Callender notion of \trie\ theories like electrodynamics and quantum mechanics do not seem to be time reversal invariant, suggesting a temporal orientation of space-time. However, the conclusion also depends on the choice of ontology. We have argued that ontologies can be considered for electrodynamics and quantum mechanics so that they are time reversal invariant. As such, whether one adopts the notion of \trie\ of Albert and Callender or the standard one, the conclusion can be the same, namely that these theories do not suggest a temporal orientation of space-time. 

We do not want to suggest that any of these ontologies are preferred. We merely wanted to point out that such ontologies do exist. An ontology respecting the relativistic character of electrodynamics remains challenging. This was only achieved with the particle-only ontology of the Wheeler-Feynman theory.

\section{Acknowledgments}
It is a pleasure to thank Craig Callender, Bryan Roberts and Sylvia Wenmackers for useful comments and discussions, and the reviewers and editors for their helpful suggestions. Support is acknowledged from the Research Foundation Flanders (Grant No. G066918N).

\end{document}